\documentclass{optica-article}

\journal{opticajournal} 

\articletype{Research Article}

\usepackage{lineno}

\begin{document}

\title{Confocal Raman Microscopy with Adaptive Optics}

\author{Juan David Muñoz-Bolaños,\authormark{1, *} Pouya Rajaeipour, \authormark{2} Kai Kummer, \authormark{4} Michaela Kress, \authormark{4} \c Ca\~glar Ataman, \authormark{2, 3} Monika Ritsch-Marte,\authormark{1} and Alexander Jesacher\authormark{1} }

\address{\authormark{1} Institute for Biomedical Physics, Medical University of Innsbruck, Müllerstraße 44, 6020 Innsbruck, Austria\\
\authormark{2}Phaseform GmbH, Georges-K\"ohler-Allee 302, 79110 Freiburg, Germany\\
\authormark{3}Microsystems for Biomedical Imaging Laboratory, Dept. of Microsystems Engineering, University of Freiburg, Georges-K\"ohler-Allee 101, 79110 Freiburg, Germany\\
\authormark{4}Institute of Physiology, Medical University of Innsbruck, Sch\"opfstraße 41, 6020 Innsbruck, Austria}

\email{\authormark{*}juan.munoz@student.i-med.ac.at} 
\begin{abstract*} 
Confocal Raman microscopy, a highly specific and label-free technique for the microscale study of thick samples, often presents difficulties due to weak Raman signals. Inhomogeneous samples introduce wavefront aberrations that further reduce these signals, requiring even longer acquisition times. In this study, we introduce adaptive optics to confocal Raman microscopy for the first time to counteract such aberrations, significantly increasing the Raman signal and image quality.  
The method is designed to integrate seamlessly with existing commercial microscopes without hardware modifications. It uses a wavefront sensorless approach to derive aberrations using an optofluidic, transmissive spatial light modulator that can be attached to the microscope nosepiece.
Our experimental results demonstrate the compensation of aberrations caused by artificial scatterers and mouse brain tissue, improving spatial resolution and achieving up to 3.5-fold  signal enhancements. Our results provide a basis for the molecular label-free study of biological systems at greater imaging depths.
\end{abstract*}

\section{Introduction}

Confocal Raman microscopy (CRM) is a label-free imaging technique that combines the chemical specificity of Raman scattering with the precise sectioning capacity of confocal microscopy~\cite{Caspers2003Combined, dieing2011confocal}. Raman signals are inherently weak, leading to lengthy acquisition times even for small-area scans.  Inhomogeneous structures of specimens introduce optical aberrations that diminish signal strength, contrast, and spatial resolution, thereby extending acquisition times even further.  
Adaptive Optics (AO) is a technology that compensates for optical aberrations and therefore restores the imaging quality. It has initially been conceived for astronomical applications \cite{babcock1953possibility, tyson2022principles}, has found routine usage in ophthalmology~\cite{Roorda2002Adaptive, akyol2021adaptive} and has been implemented in various modalities of optical microscopy ~\cite{Booth2014Adaptive}, effectively restoring image quality.

In the context of spectral microscopy, AO has been demonstrated for coherent Raman imaging~\cite{wright2007adaptive,lim2022adaptive}, surface- and tip-enhanced Raman spectroscopy~\cite{shutova2020adaptive, lee2021adaptive}, Brillouin spectroscopy~\cite{edrei2018sensor} and spectral imaging~apart from CRM \cite{wavefront2016thompson, enhanced2018Paniagua}.  Although the potential benefits of using AO in CRM have been discussed~\cite{everall2010confocal}, to the best of our knowledge this combination has remained unexplored. 
Most dynamic wavefront shapers are \emph{reflective}, which makes their integration into commercial microscopes complicated and challenging for non-experts, as major modifications to the optical path are required.  

Here we demonstrate for the first time the application of AO in CRM. Our implementation showcases the use of a \emph{refractive} wavefront shaper, a ``deformable phase plate'' (DPP)~\cite{2021JOptM1c4502R}, which can be attached to the nosepiece of a commercial microscope, making our method easily accessible even to those without extensive expertise in optics.  
We measure aberrations without a dedicated wavefront sensor, instead by using \emph{modal wavefront sensing}~\cite{Booth2003Direct}. This feedback-driven technique retrieves aberrations by displaying a series of test aberrations (``modes'') across varying magnitudes. In this context, we introduce the concept of ``spectral AO'', which denotes the procedure of extracting a "spectral guide star'' from the Raman signature to guide the optimization algorithm. Even in confocal imaging systems, the presence of guide stars is important, because it improves the performance of feedback-based AO~\cite{edrei2018adaptive}. 
Finally, we establish an approach for numerically deriving testing modes that maintain the focal point position during wavefront measurement, ensuring stable results in heterogeneous samples.  

We present experimental findings on mitigating systematic spherical aberrations caused by refractive index (RI) mismatch, as well as addressing general unknown aberrations that arise when the sample exhibits a spatially varying RI distribution. These conditions are almost always found in biological specimens that are thicker than a few cellular layers. 
We demonstrate the measurement and correction of unknown aberrations introduced by an artificial scatter and 30~µm of mouse brain tissue. 
Using AO we were able to improve Raman signals by factors of up to 3.5.  
This highlights a key benefit of applying AO in CRM: it reduces acquisition times by maximizing the signal. We anticipate that our findings will have a significant impact on research interested in label-free imaging: in biological imaging, our results could lead to the development of spectroscopic imaging that can reach deep into tissues; in the field of materials science, our findings may enable high-resolution confocal Raman imaging of challenging samples, such as structures hidden behind thick, transparent windows that cause a substantial refractive index mismatch, or multi-layered samples.

\section{Materials \& Methods}

This section outlines the materials and methods used in this study, including details on the experimental design, sample preparation, data collection, and analytical techniques. 

\subsection{Materials}

\subsubsection*{Wavefront shaper}
\begin{figure}[htpb]
    \centering
    \includegraphics[width=7 cm]{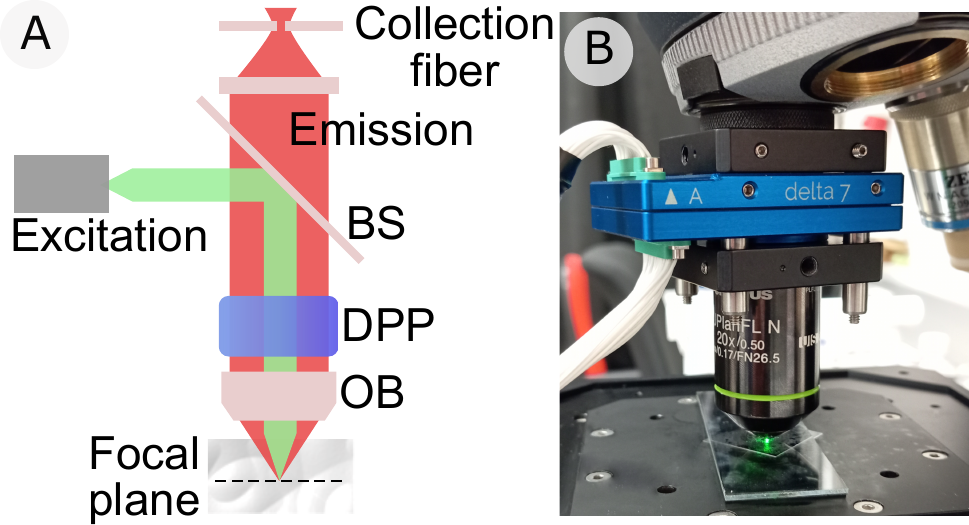}
    \caption{\textbf{Experimental setup.} (A) Optical path of the CRM. The sketch shows the excitation (green) and emission (red) beam paths, containing the DPP, objective (OB), dichroic beamsplitter (BS), and the collection fiber port. (B) Photograph of the microscope's nosepiece, with the DPP attached (blue element).}
    \label{fig:setup}
\end{figure}

The wavefront shaper is a ``Delta 7'' optofluidic deformable phase plate (DPP) from Phaseform GmbH, which contains 63 electrostatic actuators within a clear pupil of 10~mm diameter. As shown in Fig.~\ref{fig:setup}, it can be attached directly to the microscope nosepiece without any additional hardware modifications, using only a few optomechanical parts that can be purchased from common distributors. 
This particular DPP model does not have an anti-reflection (AR) coating, so its light transmission in the visible range is approximately 80\%. With AR coatings, a transmission close to 95\% should be possible. 

\subsubsection*{Imaging System}
We use a WITec alpha 300 CRM, equipped with a fiber-coupled single-mode solid-state laser emitting at 532 nm delivering an optical power of 4 mW into the objective lens, and a collection fiber (25 µm core diameter) to provide high optical sectioning. The collected Raman signal is recorded by a UHTS spectrometer (600 lines/cm)  equipped with a Newton EMCCD camera that was cooled down to -60$^\circ$C. 
To make optimal use of the DPP's wavefront shaping capability, we took care to only use objective lenses whose pupil diameters closely match the 10 mm aperture of the DPP. An objective's pupil diameter is calculated as $D_\mathrm{pupil} = 2 f_{obj} \mathrm{NA}$, where $f_\mathrm{obj}$ is its effective focal length. We use two objectives that approximately meet this condition: a dry lens from Olympus  (UPlanFl, 20x, 0.5 NA, $D_\mathrm{pupil} = 9.0~$mm) and an oil immersion lens from Zeiss (Plan-Apochromat 40x, 1.3 NA, $D_\mathrm{pupil} = 10.7$ mm).  
When utilizing a specific objective lens, the DPP applies Zernike modes to correspond to the pupil dimensions of the lens, as specified by the corresponding control files.

\subsubsection*{Mouse brain samples}
The brain samples used in this study were prepared from natural, wild-type (C57BL/6J) mouse brain tissue. 
Acute brain slices of 200~\textmu m thickness were cut on a vibrating microtome (VT1200S, Leica Microsystems, Germany), placed on microscope slides, embedded in PBS, and covered with ~100 µm coverslips (No. 0). 




\subsection{Wavefront sensing strategy}\label{sec:WFS}

\begin{figure}[htpb]
    \centering
    \includegraphics[width=10 cm]{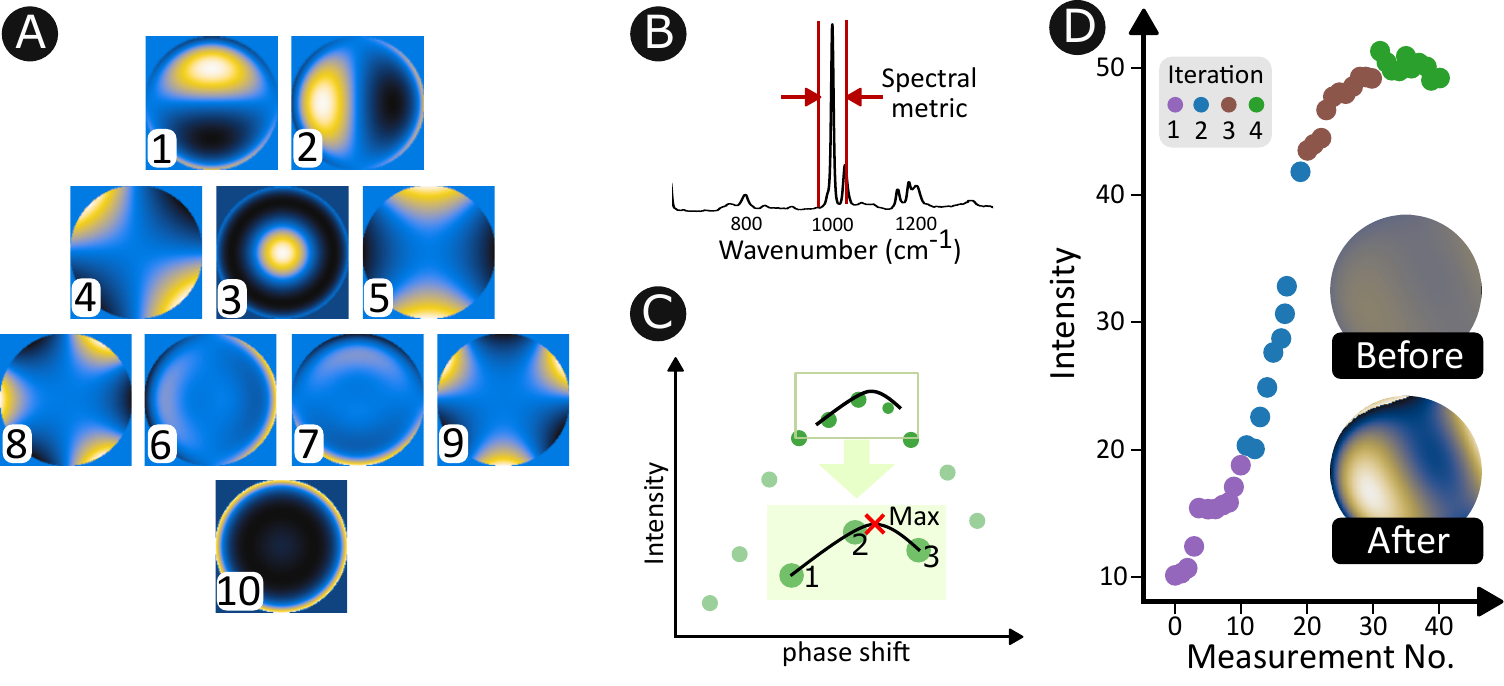}
    \caption{The spectral Adaptive Optics algorithm operates by utilizing spectral information from the sample. (A) Shift-free modes used for wavefront sensing. (B) Selecting an appropriate spectral band as metric serves as basis for optimization. (C) Displaying a specific mode at different magnitudes enables the identification of the peak signal by fitting a parabola through the peak and its two neighboring points. This process determines the optimal magnitude setting for the mode. (D) The optimization sketched in (C) is executed for each mode (defining one iteration), and the whole process is repeated up to three times (4 iterations in total).}
    
    \label{fig:algorithm}
\end{figure}

In general, there are two different approaches to measure wavefronts in microscopy: direct methods that employ wavefront sensors such as the Hartmann-Shack sensor~\cite{Tao2011Adaptive} and indirect or wavefront sensorless methods. Direct sensing using coherent backscatter has been demonstrated in confocal microscopy, but it is often problematic and known to be strongly dependent on the sample structure: the phase aberrations retrieved from specular reflections, for example, are missing geometrically odd contributions, while the magnitudes of even contributions appear doubled~\cite{Rahman2013Direct}.
Using direct wavefront sensing on the incoherent Raman signal could address this problem, but it requires the use of an extremely sensitive wavefront sensor. 
In contrast to direct methods, indirect approaches require several test measurements, which are simpler to perform experimentally. 
Here we are using modal wavefront sensing~\cite{Booth2002Adaptive}, a specific form of indirect sensing, where the phase applied within the circular aperture of the DPP is expressed as a series of  basis functions (``modes''):
\begin{equation}
    \label{eq:influence_matrix}
   \Phi(\rho,\theta)=\sum_{i=1}^{N}a_{i}X_{i}(\rho,\theta).
\end{equation}
Here, $\rho$ and $\theta$ are the normalized radial and azimuthal pupil coordinates, $N$ the number of included modes and $a_{i}$ the magnitude of the mode $X_{i}$. 
During the optimization routine, each mode is displayed in turn on the DPP with a range of diﬀerent magnitudes. For each DPP setting, a Raman spectrum is recorded and the sum intensity over the spectral peak(s) of interest is calculated. This sum spectrum represents the optimization metric $M$ to be maximized, as shown in Fig.~\ref{fig:algorithm}. 
Consequently, a set of measurements for one mode yields a sequence of values for $M$. The optimal magnitude for the mode, which maximizes $M$, can be determined by fitting a parabola using at least three points: the peak point and its adjacent points, as illustrated in Fig.~\ref{fig:algorithm}(C). 
This routine is executed for all $N$ modes, leading to a continuous increase of the Raman signal as displayed in (D). 
If the modes $X_i$ have an independent influence on the metric $M$, for instance, if $M$ can be expressed in the form $M \propto 1 - \sum_{i=0}^N c_i a_i^2$~\cite{kubby2013adaptive}, where the coefficients $c_i$ are expressing the sensitivity of $M$ to aberration modes $X_i$, full aberration compensation can theoretically be obtained already after testing each mode once. Such an orthogonality is however only achievable for the case of relatively small aberrations, in which case the aberrations can be estimated with a minimal number of $N + 1$ test measurements~\cite{booth2006wave}, although most practical implementations conduct $2N + 1$ or $3N$ measurements in total~\cite{facomprez2012accuracy}. 
For stronger aberrations, the modes may no longer remain independent for the metric $M$, requiring multiple iterations of the entire optimization process as shown in (D). Even so, typically three iterations are sufficient, even for strong aberrations. 
In this proof-of-concept study, we have included the first $N = 10$ modes (tilt, astigmatism, defocus, coma, trefoil, and spherical) and conducted 7-10 test measurements to determine each mode's peak. The entire process was then repeated two or three times, resulting in a total number of 300 measurements (i.e., spectral recordings) for the optimization. 
The time required to take a single spectral recording varied depending on the sample, ranging from 100~ms for the polymer beads in the experiments with the glass slide (Fig.~\ref{fig:spherical_proof}) and the artificial scatterer (Fig.~\ref{fig:spectral_adaptive_optics}) to 700~ms for measurements of brain tissue at 30~µm depth (Fig.~\ref{fig:brain_results}). 
Before each measurement, the DPP was allowed to settle and stabilize for 200~ms (the minimum settling time of the DPP is approximately 55 ms). 

\subsubsection*{Avoiding focal shifts during wavefront sensing}
A common problem with modal wavefront sensing is that applying a test mode can spatially shift the focus. For example, the application of coma can lead to lateral shifts, while spherical aberration modes can shift the focus axially. This effect is particularly noticeable when the excitation beam underfills the objective pupil and DPP aperture, in which case only the central region of the Zernike modes is exposed to light. 
The algorithm can be easily misled due to the variation in the focal region from which the Raman signal is collected during measurements. For instance, applying spherical test aberrations can move the focus axially to an area with a higher concentration of Raman active compounds, which results in an increased signal by adding an incorrect amount of spherical aberration. In our commercial Raman system, these shifts are significantly pronounced because the excitation laser considerably underfills the objective pupil. 
Thayil \& Booth \cite{displacementlessmodes2011} describes an effective method to create 'shift-free' modes. Their technique evaluates the image shifts induced by various aberration modes and employs a linear algebraic method to eliminate the phase components responsible for these shifts from the set of modes.
We follow essentially the same approach here, except that the derivation of a shift-free mode set is entirely computational. Detailed information on the procedure can be found in the supplementary document. 
An image of the shift-free mode set that we use for optimization throughout this work is presented in Fig.~\ref{fig:algorithm}(A).

\subsubsection*{Compensating spherical aberrations}\label{Sec: SA compensation}
The wavefront change $\Delta W$ induced by a single RI transition can be expressed in the objective pupil  as the difference of two spherical defocus functions~\cite{jesacher2010parallel}. Here, $\rho = r/(f \; \mathrm{NA})$  is the normalized radial pupil coordinate,  $r \leq f \; \mathrm{NA}$ is the radial coordinate, $f$ and $\mathrm{NA}$ are the focal length and numerical aperture of the objective and $n_1$ and $n_2$ are the RIs of the two media defining the interface. 
\begin{equation}
    \Delta W(\rho) = \left(D_{n_2}\left(\rho\right) - D_{n_1}\left(\rho\right) \right)~\Delta z,
    \label{Eq:SA}
\end{equation}
where $\Delta z$ is the imaging depth beyond the RI interface and $D_n(\rho)$ is defined as a sphere with a radius given by the refractive index $n$: 
\begin{equation}
    D_n(\rho) = \sqrt{n^2 - \rho^2~\mathrm{NA}^2}.
\end{equation}

\begin{figure}[htbp]
    \centering\includegraphics[width=10cm]{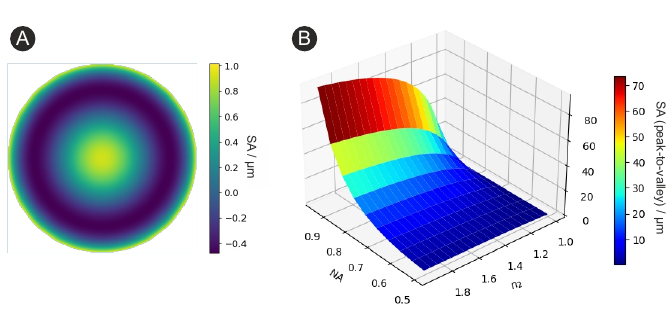}
    \caption{\textbf{The spherical aberration introduced by an RI mismatch.} (A) ``Sombrero'' part of the wavefront change $\Delta W$, representing the spherical aberration to be corrected, here shown for the experiment sketched in Fig.~\ref{fig:spherical_proof} ($\Delta z = 1$~mm, $n_1 = 1$, $n_2 = 1.52$, $\mathrm{NA} = 0.5$). (B) Peak-to-valley values of $SA(\rho)$ for different values of NA and $n_2$, assuming $\Delta z = 1$~mm and $n_1 = 1$.}
    \label{fig:sombrero}
\end{figure}

$\Delta W$ is the difference between two spheres whose radii are determined by the respective RIs. 
To correct the impact of an RI transition, it is sufficient to compensate only the $\Delta W$ deviation from a sphere. This deviation, which represents the actual spherical aberration $SA$, resembles a ``sombrero hat'' as shown in Fig.~\ref{fig:sombrero}(A). Applying $-SA$ to the DPP corrects for the spherical aberration but leaves the axial focus shift uncompensated (see section~\ref{Sec: SA compensation}). Additionally correcting for the defocus would unnecessarily increase the workload of the DPP. The spherical aberration can be obtained from Eq.~\ref{Eq:SA} by removing any contributions of spherical defocus $D_{n_2}$ from $\Delta W$~\cite{jesacher2010parallel}: 
\begin{equation}
    SA(\rho) = \Delta W - \frac{\langle\Delta W', D'_{n_2}\rangle}{\langle D'_{n_2}, D'_{n_2}\rangle}\cdot D_{n_2}.
    \label{eq: SA_pure}
\end{equation}
Here, primed quantities represent the original quantities with their mean values over the pupil area removed and $\langle ... \rangle$ defines an inner product: 
\begin{equation}
    \langle f, g\rangle = 2\int_{\rho = 0}^1 f(\rho)~g(\rho)~\rho ~d\rho
\end{equation}

\section{Results}

We evaluate the performance of AO for CRM through three experiments. The first demonstrates the compensation of spherical aberrations introduced by placing a 1~mm thick glass slab into the focusing cone. The second and third experiments describe the compensation of unknown aberrations, created by an artificial scatterer and by mouse brain tissue.  

\subsection{Compensating spherical aberrations caused by an RI mismatch}

A well-known problem in optical microscopy arises when imaging across abrupt RI transitions, such as the interface between different transparent materials~\cite{torok1997role, booth1998aberration, Diaspro2002Influence}. At such interfaces an incoming spherical wavefront becomes distorted due to the nonlinearity of Snell's law, resulting in an axial stretch of the focal spot causing degradation of the signal, image contrast and spatial resolution. 

Numerous imaging applications face the problem of an RI mismatch. For instance, in biological microscopy, tissues often possess an average RI that differs from that of the immersion media used with objective lenses such as water, glycerol or oil~\cite{pawley2006handbook}. Similarly, in material science, researchers may encounter the need to observe through glass windows, varnish layers,  multilayer polymer laminates, fluidic and semiconductor microchips ~\cite{Nakashima2004Raman, Chrimes2013Microfluidics}.  

\begin{figure}[htbp]
    \centering\includegraphics[width=7cm]{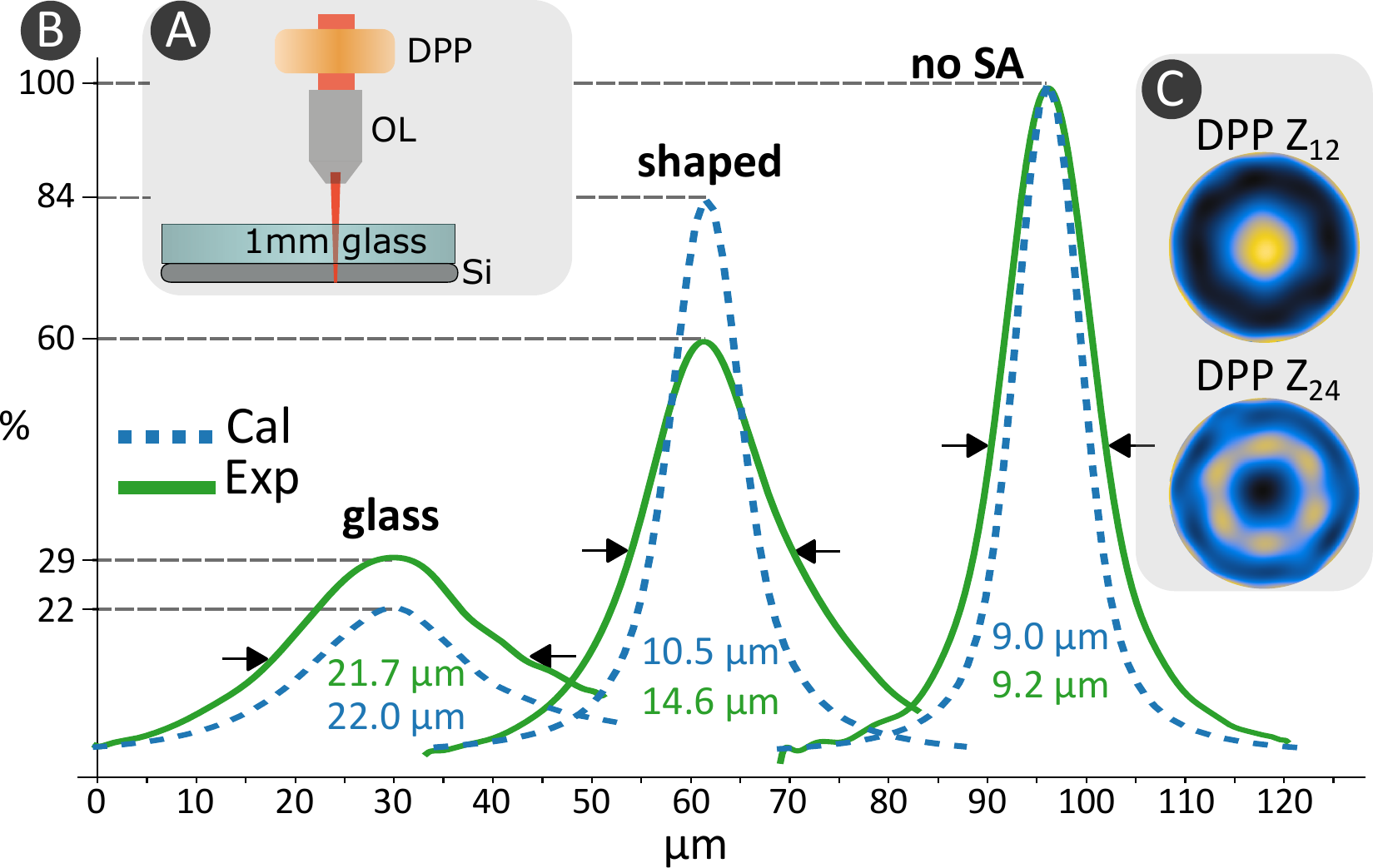}
    \caption{\textbf{Correcting spherical aberrations introduced by a 1 mm glass slide.} (A) A 1~mm thick slab of glass was placed on top of a silicon wafer chip, causing severe spherical aberrations. (B) Measured axial Raman responses for the aberrated, the corrected, and the aberration-free case. (C) Illustration of the primary 0.4 µm and 0.2 µm secondary spherical Zernike modes shaped by the DPP (calculated from the control matrix).}
    \label{fig:spherical_proof}
\end{figure}

We investigated the prospect of correcting spherical aberrations induced by an RI mismatch by introducing a 1~mm thick glass slab into the focusing cone of a 0.5NA air objective.
Because the objective is already pre-corrected for 170~\textmu m of glass and the presence of system aberrations that partially acted compensating as well, we determined the remaining glass thickness to be only around 550~\textmu m (see Supplemental Document for details). 
The presence of the glass slab predominantly shifts the focus further away from the objective lens, but also creates significant phase aberrations with an RMS value of 0.22~\textmu m.
We only correct the actual aberration (i.e., focal spot distortion), without compensating for the focal shift (see section~\ref{Sec: SA compensation}). This approach is less demanding for the DPP.
We evaluated the quality of the degraded confocal point spread function (PSF) by axially scanning across the surface of a silicon wafer fragment placed directly underneath the glass slab (see Fig.~\ref{fig:spherical_proof}(A)). 
Due to the strong absorption of silicon in the visible spectrum, we primarily detect signals originating from its surface, thus capturing the axial response of the PSF. 
The solid curves in Fig.~\ref{fig:spherical_proof}(B) represent the measured axial response of the ideal, the aberrated, and the corrected PSF, respectively. The plotted signals represent the power in the first strong Raman peak at 520 $\mathrm{cm}^{-1}$.
The dashed lines represent the matching numerical simulations. For simulating the aberration-corrected case, we calculated the wavefront shaped by the DPP using its control matrix for a 9~mm pupil. 
The ideal curve was measured with a standard coverslip (170 µm thickness No. 1.5H) on top of the silicon layer, for which the objective is designed. 
The signal was maximized using the DPP to exclude the presence of any system aberrations. 
The widths of the measured and simulated ideal response curves match well, with 9.2~\textmu m vs. 9.0~\textmu m.
Introducing the 1~mm thick glass slide results in a noticeable reduction of the peak Raman signal by a factor of 3.3 and a broadening of the response curve to 21.7~\textmu m. 
In comparison, the simulation predicts a stronger peak drop by a factor of 4.2 and a comparable width increase to 22.0~\textmu m.
The slightly broader experimental response curve, combined with the less pronounced drop at the glass insertion, indicates residual initial aberrations that could not be fully removed by the initial correction step. 

We then varied the magnitudes of primary spherical aberration modes on the DPP until the Raman response was maximal, regaining about 60\% of the ideal signal in the experiment. The response width likewise improved to 14.6~\textmu m.
These results are lower than the expected outcomes from the simulation, which forecasts a signal recovery up to 84\% of the ideal signal and a width decrease to 10.0~\textmu m.

We conclude that the DPP could only partially correct for the aberration. Further phase measurements using a Michelson interferometer indicated that, although the DPP should theoretically be capable of achieving the desired correction level, the control matrix we used was not precise enough to generate the optimal wavefront shape in an open loop. Additional details can be found in the Supplemental Document.


The correction of aberrations due to an RI mismatch does not require a wavefront measurement if the parameters in Eq.~\ref{Eq:SA} are known. Even if they are not known, the search for optimal compensation can be restricted to spherically symmetric modes, significantly reducing the overall search time.

\subsection{Addressing unknown aberrations}

Compensating for unknown aberrations such as arising from the heterogeneous structure of biological tissue requires a wavefront-sensing step. We use feedback-based wavefront sensing using shift-free Zernike modes such as described in section ~\ref{sec:WFS}. We demonstrate the measurement and compensation of unknown aberrations introduced by an artificial aberrating layer of nail polish and mouse brain tissue. 

\begin{figure}[htpb]
    \centering\includegraphics[width=7cm]{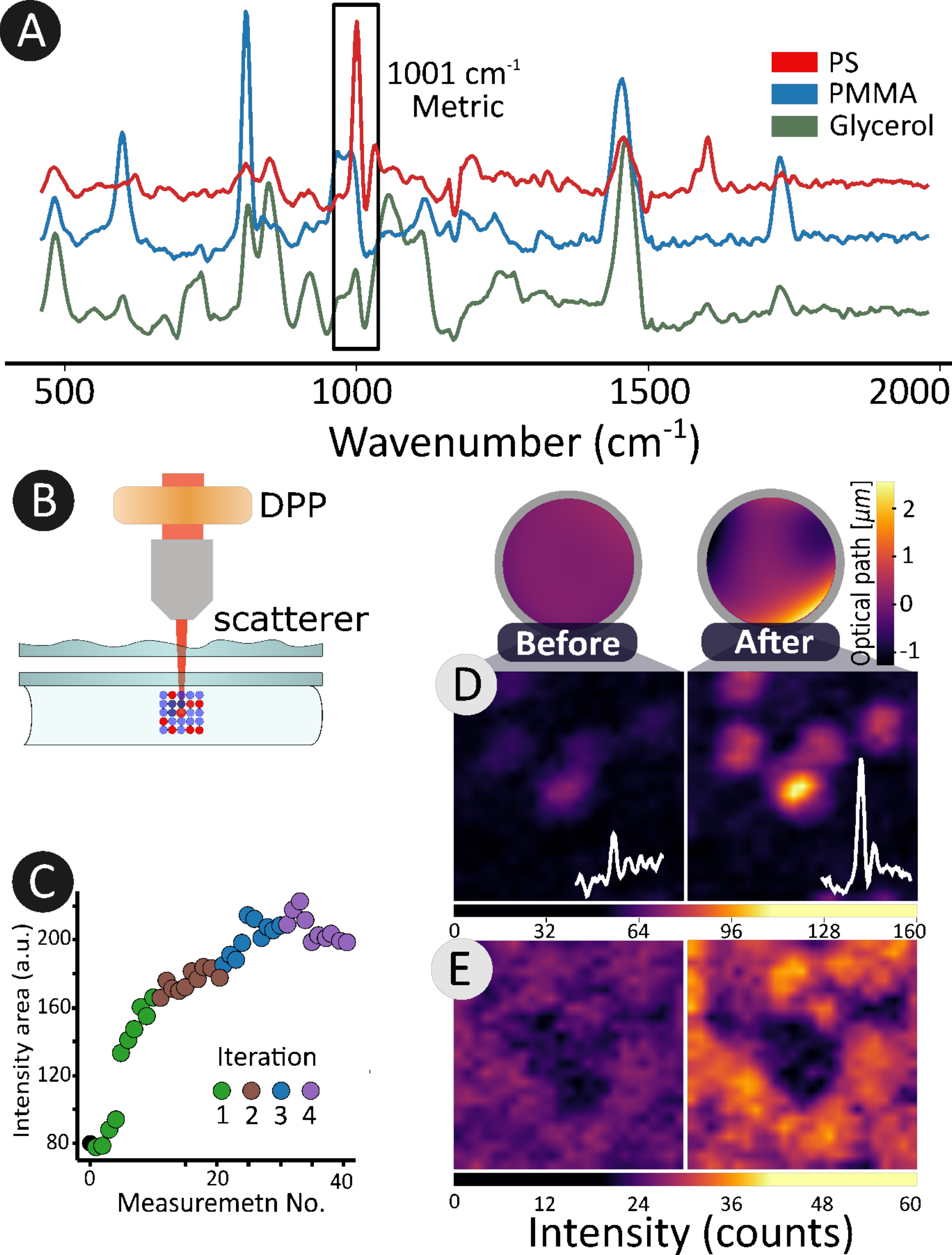}
    \caption{\textbf{Correcting unknown aberrations introduced by an artificial scatterer.} (A) The specimen contains three Raman active components in the focal region: PS, PMMA and glycerol.  The spectral band chosen for optimization is framed by the black rectangle. (B) Illustration of the sample geometry: The specimen is covered by 170~\textmu m of glass and a second cover slip featuring an aberrating nail varnish coating is placed at a distance of 1 mm. (C) Continuous improvement of the integrated signal around the PS band at 1001~$\textrm{cm}^{-1}$ during the AO routine. Over 4 iterations, a set of 10 modes was repeatedly optimized, resulting in a final signal enhancement by a factor of 2.5. (D) 2D image of PS derived from the sum signal in the interval [996, 1005] ~$\textrm{cm}^{-1}$, before (left) and after correction. The white plot shows the spectral intensity at the center of the bead, featuring a bright PS bead that served as a guide star during optimization. The circular inset images display the DPP state before (on the left) and after correction. (E) 2D image of PMMA in the same region, derived from the sum signal in the interval [596, 606] ~$\textrm{cm}^{-1}$. The colorbar of the inset images is normalized}
    \label{fig:spectral_adaptive_optics}
\end{figure}

In the first experiment, a Raman-active sample was created by mixing 10~µm polystyrene (PS) and polymethyl methacrylate (PMMA) beads with glycerol and covering the mixture with a 170~µm glass coverslip. The Raman spectra of the three components, measured using the Olympus 20x 0.5NA dry objective, are shown in Fig.~\ref{fig:spectral_adaptive_optics}(A). 
A second coverslip was placed approximately 1 mm above the sample, coated on its upper surface with an aberrating thin layer of transparent nail varnish (B). 

We selected the PS band around 1001~$\mathrm{cm}^{-1}$ to be optimized. This specific spectral selection effectively transformed a single PS bead in the focal region into a ``guide star'' that stands out of the background made from other substances such as glycerol and PMMA.
In wavefront sensing, the presence of guide stars is often crucial for the optimization process. They act as small beacons defining the wavefront to correct. In Raman microscopy,  specific regions or particles inside the specimen can be isolated by selecting their respective spectral bands for optimization. This method allows for the utilization of data from Raman spectra in wavefront sensing, which ensures effective aberration measurements even without a strict confocal gate. This is particularly useful when larger collection fibers are employed to enhance the signal or in-line scanning systems. 
In this particular experiment, the iterative wavefront sensing and correction routine as outlined in section~\ref{sec:WFS} led to a 2.5-fold increase in the Raman signal. The continuous signal improvement during the optimisation process is illustrated in Fig.~\ref{fig:spectral_adaptive_optics}(C).

Figures~\ref{fig:spectral_adaptive_optics}(D) and (E) illustrate 2D Raman images captured before and after correction, respectively.  These images are constructed from the integrated signal in the PS (D) and PMMA (E) bands as stated in the figure caption. 

\begin{figure}[htpb]
    \centering\includegraphics[width=7cm]{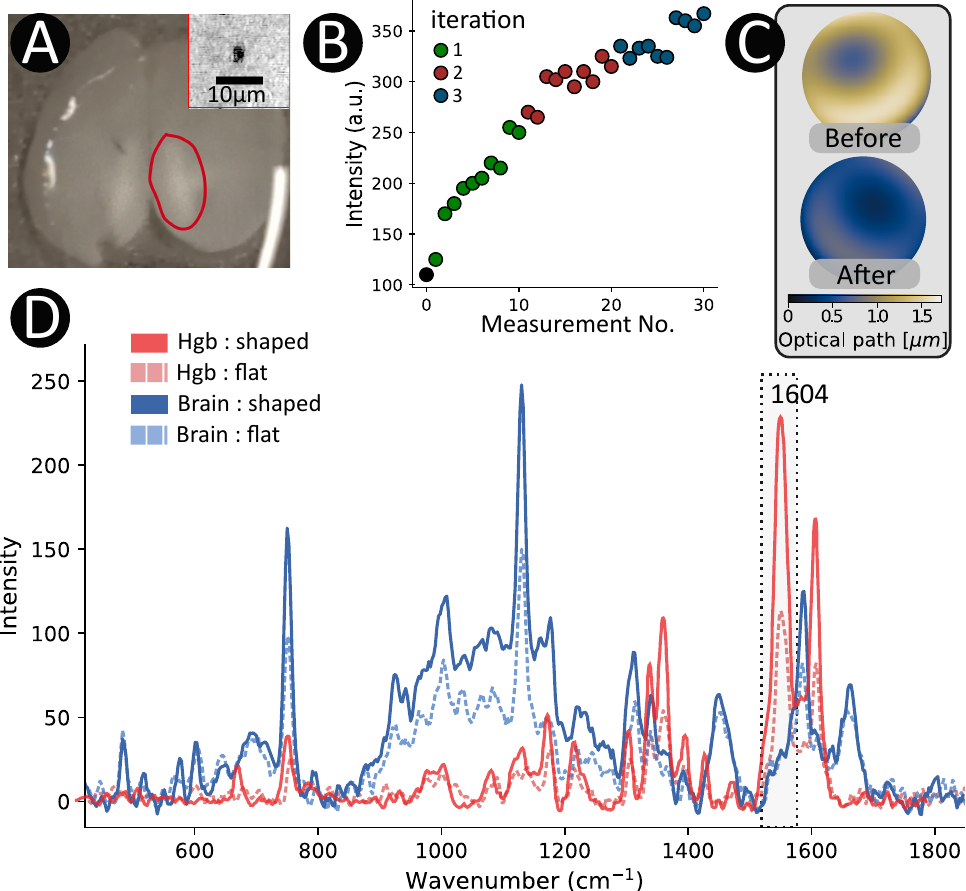}
    \caption{\textbf{Correction of tissue aberrations} (A) Microphoto of the mouse brain slice, with the Raman imaged region marked in red. The upper gray widefield microscope image shows a black spot, which is the single erythrocyte at 30~µm depth that we used as guide star. (B) Performance of the optimization algorithm over 3 iterations for 10 modes. (C) DPP shapes before and after correction, shown as phase images. (D) Raman spectra of hemoglobin and brain tissue before and after correction. }
    \label{fig:brain_results}
\end{figure}

We investigate the measurement and correction of aberrations occurring when imaging inside mouse brain tissue. Figure~\ref{fig:brain_results} illustrates the outcomes of AO applied to brain imaging using a 40x oil immersion objective (1.3NA). In this experiment, the comparatively strong Raman signal from hemoglobin (Hgb) within a single erythrocyte at a depth of 30~µm within the brain tissue worked as a spectral guide star. As a consequence of the optimization process, a small ``window'' was opened around the position of the red blood cell (RBC), which in turn increased the Raman signals from the surrounding brain tissue as well. 
The total time required for scattering brain optimization (10 modes corrected in 3 iterations with 9 increasing phase steps) amounted to \~8 min regarding Raman acquisition (700 ms for two measurements) and DPP configuration (200 ms). 

Fig.~\ref{fig:brain_results}(D) illustrates the spectral signatures of deoxygenated hemoglobin with its characteristic bands at 1604 (C=C stretching within porphyrin breathing), 1550 (symmetric stretching of the porphyrin ring), and 1356 (Fe-H stretch) $\mathrm{cm}^{-1}$ ~\cite{scientific132023}  and brain tissue with characteristic bands at 1655 (Amide I), 1582 (C=C stretching of Phenylalanine), 1440 (C-H bending),  1150 (C-C and C-N stretching, lipids/proteins), 1001 (symmetric ring breathing of Phenylalanine), 915 (glycogen and carbohydrates, C-C stretching) and 745 (out-of-plane bending - nucleic acids - DNA/RNA) $\mathrm{cm}^{-1}$~\cite{cancers13050960, bios13010027, Abrahmczyk2021Redox}.  The spectral band around 1550 $\mathrm{cm}^{-1}$ was chosen to optimize the signal, resulting in a 2.2-fold increase of the Hgb signal. 
As mentioned above, aberration correction increases the signal not only at the position of the guide star, but also at nearby imaging points. The size of the corrected patch depends on various parameters such as the scattering properties of the tissue, the imaging depth and the wavelength. 
We took a Raman measurement of brain tissue at a distance of 5 µm to the RBC, where we could still observe an around 2-fold signal improvement. This demonstrates the applicability of the spectral guide star approach to Raman imaging, where signal from the tissue of interest (the brain cells) can be enhanced by optimizing the signal from an adjacent, strong beacon, such as the red blood cell.

\section{Discussion \& Conclusion}

We demonstrate, to our knowledge, the first application of AO in confocal Raman microscopy. 
The technique compensates for the degrading effect of aberration-introducing samples on both the incident laser wavefront and the epi-detected Raman light, which can significantly improve signal, contrast, and spatial resolution of confocal Raman images. 
We have experimentally demonstrated the partial compensation of spherical aberrations caused by an RI mismatch, as well as the correction of unknown aberrations introduced by an artificial scatterer and by mouse brain tissue. Our results show up to 3.5-fold signal improvements, which enable shorter acquisition times without sacrificing signal-to-noise ratio. These shorter recording times are one of the primary motivations for the use of AO in CRM.

Our particular approach using a transmissive wavefront shaper is appealing to those without expertise in optics because it can be easily attached to the nosepiece of a commercial microscope. Additionally, we employ a straightforward wavefront sensorless measurement approach to retrieve aberration information.

We show how the abundance of information in Raman spectra can be used to define ``spectral guide stars'' within the sample by selecting appropriate spectral bands to define the optimization metric. These guide stars, in addition to the gating effect of the pinhole, ensure a monotonic relationship between the aberration magnitude and metric, which is a necessary condition for the hill-climbing optimization routine. Successful aberration measurements can even be achieved when using larger collection fibers (i.e. larger pinholes) or in Raman line scanning systems, which generally have weaker axial sectioning capability. 

We note that the richness of spectral information also enables the definition of more efficient optimization metrics. For instance, it would be possible to define the metric based on the outcome of a cluster analysis or to utilize the power ratio of Raman peaks belonging to the guide star and the surrounding medium. More efficient metrics would show a stronger, more clear dependence on aberrations, therefore facilitating shorter wavefront sensing times. 

When combining indirect wavefront sensing with Raman signals it is important to minimize the measurement time. Our procedure has not yet been optimized for speed, so there is room for improvement. For example, if the dependence of the metric on the magnitude of the mode is weak, as shown in Fig.~\ref{fig:algorithm}(C), fewer measurements are sufficient to determine the peak position. Similarly, the estimated magnitude of an aberration mode typically varies little after the first measurement iteration, allowing the number of measurements required to be reduced from 10 (as used here) to just 3, the minimum number required to fit a parabola to the data.
In optimal cases, the number of spectral measurements required to correct for 10 modes could therefore be reduced from the 300 used here to just 21 (following the ``2N+1'' scheme), speeding up the process by a factor of more than 10.  
Nevertheless, we believe that direct aberration measurements, such as those obtained using a Hartmann-Shack or pyramid wavefront sensor, would significantly improve the feasibility of AO in confocal Raman microscopy. However, these techniques introduce additional complexity and are not straightforwardly compatible with the definition of spectral guide stars as introduced here.



\begin{backmatter}
\bmsection{Funding}
We gratefully acknowledge funding from the FWF (P32146-N36) and FWF (DOC. 110-B). 

\bmsection{Data availability} Data underlying the results presented in this paper are not publicly available at this time but may be obtained from the authors upon reasonable request.







\end{backmatter}

\bibliography{sample}






\end{document}